\documentstyle[prl,aps]{revtex}

\begin{document}
\preprint{DRAFT-- SLAC-PUB-96-7242, \today (T/E)}

\title{Measurement of the Proton and Deuteron Spin
Structure Function $g_1$
in the Resonance Region}

\author{
K.~Abe,$^{15}$
T.~Akagi,$^{12,15}$
P.~L.~Anthony,$^{12}$
R.~Antonov,$^{11}$
R.~G.~Arnold,$^{1}$
T.~Averett,$^{16,\ddag\ddag}$
H.~R.~Band,$^{18}$
J.~M.~Bauer,$^{7}$
H.~Borel,$^{5}$
P.~E.~Bosted,$^{1}$
V.~Breton,$^{3}$
J.~Button-Shafer,$^{7}$
J.~P.~Chen,$^{16,8}$
T.~E.~Chupp,$^{8}$
J.~Clendenin,$^{12}$
C.~Comptour,$^{3}$
K.~P.~Coulter,$^{8}$
G.~Court,$^{12,*}$
D.~Crabb,$^{16}$
M.~Daoudi,$^{12}$
D.~Day,$^{16}$
F.~S.~Dietrich,$^{6}$
J.~Dunne,$^{1}$
H.~Dutz,$^{12,**}$
R.~Erbacher,$^{12,13}$
J.~Fellbaum,$^{1}$
A.~Feltham,$^{2}$
H.~Fonvieille,$^{3}$
E.~Frlez,$^{16}$
D.~Garvey,$^{9}$
R.~Gearhart,$^{12}$
J.~Gomez,$^{4}$
P.~Grenier,$^{5}$
K.~A.~Griffioen,$^{11,17}$
S.~Hoibraten,$^{16,\S}$
E.~W.~Hughes,$^{12,\ddag\ddag}$
C.~E.~Hyde-Wright,$^{10}$
J.~R.~Johnson,$^{18}$
D.~Kawall,$^{13}$
A.~Klein,$^{10}$
S.~E.~Kuhn,$^{10}$
M.~Kuriki,$^{15}$
R.~Lindgren,$^{16}$
T.~J.~Liu,$^{16}$
R.~M.~Lombard-Nelsen,$^{5}$
J.~Marroncle,$^{5}$
T.~Maruyama,$^{12}$
X.~K.~Maruyama,$^{9}$
J.~McCarthy,$^{16}$
W.~Meyer,$^{12,**}$
Z.-E.~Meziani,$^{13,14}$
R.~Minehart,$^{16}$
J.~Mitchell,$^{4}$
J.~Morgenstern,$^{5}$
G.~G.~Petratos,$^{12,\ddag}$
R.~Pitthan,$^{12}$
D.~Pocanic,$^{16}$
C.~Prescott,$^{12}$
R.~Prepost,$^{18}$
P.~Raines,$^{11}$
B.~A.~Raue,$^{10,\dag}$
D.~Reyna,$^{1}$
A.~Rijllart,$^{12,\dag\dag}$
Y.~Roblin,$^{3}$
L.~S.~Rochester,$^{12}$
S.~E.~Rock,$^{1}$
O.~A.~Rondon,$^{16}$
I.~Sick,$^{2}$
L.~C.~Smith,$^{16}$
T.~B.~Smith,$^{8}$
M.~Spengos,$^{1,11}$
F.~Staley,$^{5}$
P.~Steiner,$^{2}$
S.~St.Lorant,$^{12}$
L.~M.~Stuart,$^{12}$
F.~Suekane,$^{15}$
Z.~M.~Szalata,$^{1}$
H.~Tang,$^{12}$
Y.~Terrien,$^{5}$
T.~Usher,$^{12}$
D.~Walz,$^{12}$
F.~Wesselmann,$^{10}$
J.~L.~White,$^{1,12}$
K.~Witte,$^{12}$
C.~C.~Young,$^{12}$
B.~Youngman,$^{12}$
H.~Yuta,$^{15}$
G.~Zapalac,$^{18}$
B.~Zihlmann,$^{2}$
D.~Zimmermann$^{16}$}

\address{
{\rm (E143 Collaboration)}\break
{$^{1}$The American University, Washington, D.C. 20016}  \break
{$^{2}$Institut f\" ur Physik der Universit\" at Basel,
  CH--4056 Basel, Switzerland} \break
{$^{3}$LPC IN2P3/CNRS,
  University Blaise Pascal, F--63170 Aubiere Cedex, France}  \break
{$^{4}$TJNAF, Newport News, Virginia 23606} \break
{$^{5}$DAPNIA-Service de Physique Nucleaire
  Centre d'Etudes de Saclay, F--91191 Gif/Yvette, France} \break
{$^{6}$Lawrence Livermore National Laboratory, Livermore, California 94550}
\break
{$^{7}$University of Massachusetts,  Amherst, Massachusetts 01003}  \break
{$^{8}$University of Michigan, Ann Arbor, Michigan 48109} \break
{$^{9}$Naval Postgraduate School, Monterey, California 93943} \break
{$^{10}$Old Dominion University,  Norfolk, Virginia 23529} \break
{$^{11}$University of Pennsylvania,  Philadelphia, Pennsylvania
  19104} \break
{$^{12}$Stanford Linear Accelerator Center,
  Stanford, California 94309} \break
{$^{13}$Stanford University, Stanford, California 94305} \break
{$^{14}$Temple University, Philadelphia, Pennsylvania 19122}  \break
{$^{15}$Tohoku University, Sendai 980, Japan} \break
{$^{16}$University of Virginia, Charlottesville, Virginia 22901} \break
{$^{17}$The College of William and Mary, Williamsburg, Virginia 23187} \break
{$^{18}$University of Wisconsin, Madison, Wisconsin 53706}  \break
{\rm (Accepted in Phys. Rev. Lett.)}
}
\maketitle


\begin{abstract}
We have measured the proton and deuteron spin
structure functions $g_1^p$ and $g_1^d$ in the region of the
nucleon resonances for $W^2 < 5$ GeV$^2$ and
$Q^2\simeq 0.5$ and $Q^2\simeq 1.2$ GeV$^2$ by inelastically scattering
9.7 GeV polarized electrons off polarized $^{15}$NH$_3$ and
$^{15}$ND$_3$ targets. We observe significant structure in $g_1^p$ in
the resonance region.  We have used
the present results, together
with the deep-inelastic data at higher $W^2$, to extract
$\Gamma(Q^2)\equiv\int_0^1 g_1(x,Q^2) dx$.  This is the first information on the low-$Q^2$
evolution of $\Gamma$ toward the Gerasimov-Drell-Hearn
limit at $Q^2 = 0$.


\end{abstract}


\narrowtext
\twocolumn

The nucleon spin structure functions $g_1$ and $g_2$
have been and continue to be intensively studied via
deep-inelastic lepton scattering\cite{E80,EMC,E142,SMC,E143,e143q};
however, they
remain largely undetermined in the resonance region, where the
squared invariant mass of the final state $W^2$ is less than 4 GeV$^2$.
Here, $g_1$ and $g_2$ can be used to investigate the helicity structure of the resonance
transition amplitudes.
States of definite spin and parity are more easily understood
in terms of the virtual photon asymmetries\cite{roberts}
\begin{eqnarray}
A_1(x,Q^2) &= {\displaystyle {\sigma_{1/2} - \sigma_{3/2}}\over
{\displaystyle \sigma_{1/2} + \sigma_{3/2}}} = {\displaystyle 1\over\displaystyle{F_1}}
\left[ g_1 - {\displaystyle Q^2\over{\displaystyle\nu^2}}g_2\right]\quad {\rm and}\nonumber\\
A_2(x,Q^2) &= {\displaystyle\sigma_{LT}\over{\displaystyle\sigma_{T}}}
= {{\displaystyle \sqrt{Q^2}}\over{\displaystyle{\nu F_1}}}
\left[g_1 + g_2\right],
\end{eqnarray}
in which $Q^2$ is the squared four-momentum transfer, $\nu$ is the energy
transfer, $x = Q^2/2M\nu$, $M$ is the nucleon mass, and the structure functions
$F_1$, $g_1$ and $g_2$ depend on both $x$ (or $W^2$) and $Q^2$.
The cross sections $\sigma_{1/2}$
and $\sigma_{3/2}$ measure
the strength of 
virtual transverse
photon absorption leading to
final-state spin projections of
$\frac{1}{2}$ and  $\frac{3}{2}$.  The cross sections $\sigma_L$,
$\sigma_T\equiv (\sigma_{1/2}+\sigma_{3/2})/2$, and $\sigma_{LT}$
measure longitudinal, transverse and interference photon absorptions.
Positivity limits require that $|A_1|\le 1$ and $|A_2| \le 
\sqrt{R(x,Q^2)}$, in which $R \equiv {\sigma_L/\sigma_T}$.
The excitation of
the $\Delta(1232)$ resonance (spin-$\frac{3}{2}$) allows for both $\frac{1}{2}$
and $\frac{3}{2}$ spin projections, and at low $Q^2$ is expected to be
primarily a  magnetic dipole transition, for which
$\sigma_{3/2}/\sigma_{1/2} = 3$ and $A_1 = -\frac{1}{2}$.  On the other hand,
the $S_{11}(1535)$ resonance has no spin-$\frac{3}{2}$ projection,
so $A_1$ should be unity.
The observed values of $A_1$ in the resonance region are a combination
of asymmetries for individual resonances and for the nonresonant background.
The goals of the present measurements are to gain
a better understanding of the resonances and to
determine their influence
on the deep-inelastic results, both in terms of radiative corrections and
the evolution of $\Gamma(Q^2)\equiv\int_0^1 g_1(x,Q^2) dx$.
Although the resonant contribution to $\Gamma$ is
insignificant at high $Q^2$, it
dominates the integral below about $Q^2 = 0.5$ GeV$^2$.  In fact, the limits set by the
Gerasimov-Drell-Hearn (GDH) sum rule\cite{GDH} indicate that $\Gamma(Q^2)$
should change
sign in the region $0 < Q^2 < 1$ GeV$^2$ 
and approach zero as 
$Q^2 \rightarrow 0$. 


The data for the present analysis are part of the E143 data 
set\cite{E143,e143q} taken with a 9.7 GeV polarized
electron beam (average polarization 85\%) and cryogenic
$^{15}$NH$_3$ and $^{15}$ND$_3$ targets (average polarizations of
65\% and 25\%, respectively). The data taken with the two spectrometers
at 4.5$^\circ$ and 7$^\circ$ corresponded to $Q^2 \simeq 0.5$ and $1.2$ GeV$^2$
in the resonance region. 

Since these data were taken with longitudinal target
polarization only, the determination of $g_1$ or
$A_1$ requires additional assumptions about
either $g_2$ or $A_2$. We extract $g_1$ rather than $A_1$ because
$g_1$ is significantly less affected by
our lack of knowledge of $g_2$ or $A_2$.
We set $A_2 = 0$ (corresponding to $g_1 = - g_2$)
in our analysis. This is motivated by the fact that $|A_2| \le \sqrt{R}$
and existing data\cite{keppel} indicate that $R$ is small in the
resonance region ($R=0.06\pm 0.02$ for $1<Q^2<8$ GeV$^2$ and $W^2<3$ GeV$^2$).
We have explored the sensitivity to $A_2$
by considering the alternate possibilities $g_2 = 0$ and
$g_2 = g_2^{\rm WW} = -g_1 + \int_x^1 g_1(x^\prime)
/x^\prime dx^\prime$,
the Wandzura-Wilczek\cite{wand} twist-two
form.  Maximum deviations in $g_1$ from the $A_2=0$ case
define the systematic errors due to uncertainty in $A_2$. Even if $A_2$ were
as large as 0.3, the extracted values of $g_1$ would shift by less than 0.014,
which is small compared to the statistical errors on each point.

We have extracted  $g_1$ from the absolute
cross section differences for electron and nucleon spins aligned along the
beam axis either parallel ($\uparrow\uparrow$) or antiparallel ($\uparrow\downarrow$)
to each other\cite{roberts}:
\begin{eqnarray}
{\displaystyle 1\over{\displaystyle\sigma_{\rm M}}}&
{\displaystyle d\sigma^{\uparrow\downarrow(\uparrow\uparrow)}\over
{\displaystyle d\Omega dE^\prime}}
= {\displaystyle F_2\over{\displaystyle\nu}} + {\displaystyle 2\over {\displaystyle M}}
\tan^2 {\left(\theta\over 2\right)} F_1
+(-)  {\displaystyle 2\over{\displaystyle M\nu}}
\tan^2{\left(\theta\over 2\right)} \nonumber\\
\times & \left[ (E+E^\prime\cos\theta+ Q^2/\nu) g_1 -
\sqrt{Q^2} F_1 A_2\right]
\end{eqnarray}
in which $\sigma_M$ is the Mott scattering cross section,
$E$ ($E^\prime$) is the initial (final) electron energy,
$\theta$ is the laboratory scattering angle, $\nu=E-E^\prime$, and
$F_1$ and $F_2$ are the unpolarized structure functions.

This method requires good knowledge of
spectrometer acceptances, the number
density of polarizable protons or deuterons 
in the target, and detector efficiencies. 
Alternatively, one could extract  $g_1$ from the count rate
asymmetry as in Ref.~\cite{E143,e143q}; however,
the dilution factor (the fraction of scatterings coming from a polarizable
nucleon in the target) is needed in this case, which
is more difficult to obtain reliably in the
resonance region.  Nevertheless, when we tried both methods, we found
that they agreed to within a fraction of the statistical errors on each point
(typically better than 3\%).

As the first step to determine the absolute cross section differences, we
calculated the raw count
differences per incident charge
$N^{\uparrow\downarrow}/q^{\uparrow\downarrow} - N^{\uparrow\uparrow}
/q^{\uparrow\uparrow}$, with each term corrected
for dead-time in the trigger electronics.  For each electron,
$W^2$ was calculated using
the momentum and scattering angle determined from tracking.
The data were binned in $W^2$, normalized by the product of target and beam
polarizations and corrected for absolute spectrometer efficiency.
Each detector's efficiency was determined by making a strict cut to select
good electron events without
using one of the detectors and checking how often that detector registered the
electron.  The absolute spectrometer
efficiency is the product of all of the individual detector
efficiencies (no evidence for correlations between them was found)\cite{raines}.

Fully corrected cross section differences $d\Delta\sigma/dE^\prime d\Omega$
were obtained with the help of a Monte Carlo simulation.
This simulation was used to normalize the raw data for target
density and spectrometer acceptance.  The normalized data were
then corrected for radiative and resolution effects by an
additive term determined from the Monte Carlo routine.
Small corrections for polarized
$^{15}$N and $^{14}$N in the target (and $^1$H in the case of the deuteron)
were applied as in Ref.~\cite{E143}.

The Monte Carlo code  simulated all 
relevant aspects of the
experiment and was able to predict total count rates and count 
rate differences based
on a set of input tables of cross sections and asymmetries.
The unpolarized cross sections were calculated
from a fit by Stuart {\it et al.}\cite{stuart}.
The asymmetry tables were calculated from a combination of
resonant and non-resonant contributions. The non-resonant part came
from a parameterization of all existing  deep inelastic data (Fit III of
Ref.~\cite{e143q}), which was extrapolated into the resonance region. 
The asymmetry of the $\Delta(1232)$ resonance was fixed at $A_1=-0.5$.
The contribution to the asymmetry from all other resonances was
approximated by two constant values (one for the region
below $W^2= 2.6$ GeV$^2$ and the other for the region above
$W^2 = 3.2$ GeV$^2$) and a linear interpolation between them.
This simple parameterization
was sufficient to achieve a reasonable fit to the data for the purpose
of radiative and smearing corrections.  The quality of the Monte Carlo
results was also tested by comparing the predicted and measured total
count rates and (quasi-) elastic asymmetries.  We found good agreement
within the statistical uncertainty of the data.


The full model was used to calculate cross sections and asymmetries
both in lowest order of QED (Born approximation) and with full radiative
corrections following the prescription by Shumeiko {\it et al.}\cite{radcor}.
We used the peaking approximation after convincing ourselves that the full code
produced negligible differences. The radiated cross sections and asymmetries
were tabulated as input for the Monte Carlo code, while the Born cross
sections and asymmetries were compared with the Monte Carlo output to
determine the normalization factor and the additive correction for the raw
data.



Fig.~1 shows $g_1$ obtained from
cross section differences for proton and deuteron (per nucleon)
measured with the two spectrometers as a function of $W^2$.
The full length of the error bars corresponds to the statistical and systematic
uncertainties added in quadrature. The cross bars indicate statistical
errors alone, which
dominate the total errors.
Plotted as triangles are the data
of Baum {\it et al.}\cite{baum} taken at similar
kinematics and converted to $g_1$ for comparison by assuming
$A_2=0$.
Within errors, the two measurements agree well.  
The solid lines show the Monte Carlo simulation.
The dashed curve is
a calculation using the code AO by
Burkert and Li\cite{burkert}.  
This calculation does not adequately describe the data above $W^2 = 3$
GeV$^2$, because the non-resonant background contains only the single
$\pi$ Born term.  Although AO describes the resonance region at higher
$Q^2$ rather well, there is a significant discrepancy in the second
resonance region ($2 < W^2 < 3$ GeV$^2$) at the lower $Q^2$.

Extracting the total (resonant plus non-resonant)
$A_1$ at fixed $W^2$ from the relationship
$g_1 = {\nu^2\over{\nu^2+Q^2}}( A_1+{Q\over \nu} A_2) F_1$
requires several simplifying assumptions.
We assume 
$A_2=0$, $R=0.25/Q^2$ for the non-resonant cross
section\cite{stuart}, and $R=0$ for the resonant cross section\cite{keppel}
and obtain $F_1$ from measured values of $F_2$ and $R$.  
In that case, all data for both proton
and deuteron are consistent with $A_1^\Delta = -{1\over2}$ with a relative
uncertainty of 40-100\%.  This is also the asymmetry that AO predicts.
On the other hand, $A_1$ in the region $W^2 = 2-3$ GeV$^2$, 
which is dominated by
the $S_{11}$ and $D_{13}$ resonances, is surprisingly high.  Here the proton
data at $4.5^\circ$ and $7^\circ$ are consistent with $A_1 = 0.9\pm 0.2$,
which is close to the positivity limit and well above the AO prediction.
This could indicate a significant contribution from $A_2$, from a
non-resonant asymmetry that is much larger than expected from deep-inelastic
extrapolations, or from a stronger than expected contribution from $S_{11}$
for which $A_1=1$.  The smaller corresponding asymmetry for the deuteron
($A_1 = 0.3\pm 0.2$) may arise from Fermi smearing or a genuine n-p
difference.  More detailed information on the asymmetries of individual 
resonances and the non-resonant background are expected once
high-precision, semi-exclusive data from TJNAF become available.

Fig.~2 shows the integrals $\Gamma(Q^2)$ for proton and neutron,  
evaluated at the average $Q^2$ for the resonance region ($M^2<W^2<4$ GeV$^2$).
We summed our resonance results directly 
(where $Q^2$ does not vary much) and then added 
a contribution from smaller
$x$ (larger $W^2$) at the same fixed $Q^2$
taken from Fit III to the world's deep-inelastic $g_1$ data\cite{e143q}.
The neutron
integrals were derived assuming a 5\% D-state probability for the
deuteron.  The statistical errors assigned to Fit III\cite{e143q} at given
values of $x$ and $Q^2$
corresponded to the kinematically closest E143 data points 
 at 9.7 or 16 GeV,
which dominated the fit in this region.
Systematic errors
were calculated using the systematic uncertainties for the measured $g_1$ in
the resonance region added linearly to the systematic errors for the deep-inelastic
region, which are highly correlated with each other.
Extrapolation errors for the region below the last measured datum at
$x=0.03$ were taken to be as large as the value that
Fit III yields for $x<0.03$.

Although several models for the $Q^2$ evolution of $\Gamma(Q^2)$
exist\cite{BKM,Ji,LiLi,BuIo,soffer}, we show here
only two representative ones, together with the evolution\cite{larin} of the world's
deep-inelastic data due to the changing coupling constant $\alpha_S$.
Although the GDH sum rule is strictly valid
only at $Q^2=0$ where
$\Gamma(Q^2)$ vanishes, it can be used to predict the slope of $\Gamma(Q^2)$ for
small $Q^2$.  The solid line at low $Q^2$ in Fig.~2 shows $\Gamma = -\kappa^2Q^2/8M^2$
in which $\kappa$ is the anomalous magnetic moment of either the proton or
neutron.  
Burkert and Ioffe\cite{BuIo} consider the contributions from the resonances
using the code AO, and the nonresonant contributions 
 using a simple higher-twist-type form fitted to the deep-inelastic
data.  Their model is constrained to fit both the GDH and the deep-inelastic limits,
and it describes the data quite well.
Soffer and Teryaev\cite{soffer}
assume that the integral over $g_1 + g_2$ varies smoothly from
high $Q^2$ where $g_2\approx 0$ down to $Q^2=0$.  Using
their simple prediction for this integral and subtracting the contribution
from $g_2$ alone using the Burkhardt-Cottingham sum rule\cite{buco},
gives the dashed curve in Fig.~2b,
which also agrees quite well with our data.

The present spin structure function data in the region of the nucleon resonances
allow us to determine the integrals $\Gamma(Q^2)$ for the first time at $Q^2$ below
2 GeV$^2$.  In contrast to the nearly flat behavior in the deep-inelastic region above
$Q^2=2$ GeV$^2$, $\Gamma$ varies rapidly below $Q^2=2$ GeV$^2$.
Models that interpolate between the deep-inelastic and GDH limits
describe the data quite well in this non-perturbative regime.

This work was
supported by the Department of Energy; the National Science Foundation;
the Schweizersche Nationalfonds; the Commonwealth of
Virginia; the Centre National de la Recherche Scientifique
and the Commissariat a l'Energie Atomique (French groups);
the Japanese Ministry of Education, Science, and Culture; and the Jeffress
Memorial Trust (W\&M).

\begin{table}[t]
\caption{Integrals $\Gamma(Q^2)$ of the structure functions
$g_1$ for the proton (p), deuteron (d) and neutron (n).  The
measured sum $\Gamma_{res}$ for the  resonance region ($W^2<4$ GeV$^2$)
is listed separately from the totals $\Gamma_{tot}$, which
includes the deep-inelastic region as given by fits to the
world's data.}
\begin{tabular}{ccrr}
$Q^2$ (GeV$^2$)& ~ &$\Gamma_{\rm res}$ ($\pm$stat.$\pm$syst.)&$\Gamma_{\rm tot}$($\pm$stat.$\pm$syst.)\\
\hline
0.5&p&0.026$\pm$0.008$\pm$0.008&$0.049\pm 0.008\pm 0.013$\\
0.5&d&0.000$\pm$0.013$\pm$0.008&$0.003\pm 0.013\pm 0.010$\\
0.5&n&-0.026$\pm$0.028$\pm$0.020&$-0.043\pm 0.029\pm 0.025$\\
\hline
1.2&p&0.040$\pm$0.003$\pm$0.004&$0.100\pm 0.005\pm 0.012$\\
1.2&d&0.026$\pm$0.006$\pm$0.004&$0.043\pm 0.008\pm 0.007$\\
1.2&n&0.016$\pm$0.013$\pm$0.010&$-0.006\pm 0.019\pm 0.020$\\
\end{tabular}
\end{table}
\eject

\begin{table}[t]
\caption{Fully corrected cross section differences
$d(\sigma^{\uparrow\downarrow}-\sigma^{\uparrow\uparrow})/dE^\prime
d\Omega\equiv \Delta\sigma$ (nb/GeV-sr) and structure
functions $g_1$ for the proton in the resonance region.
Listed are the data for the $4.5^\circ$ and $7.0^\circ$
spectrometers.  The values of $Q^2$ and $W^2$ (GeV$^2$) are given at
bin centers.  The additive correction $\Delta\sigma_R$ includes
both radiative effects and resolution corrections.}
\begin{tabular}{rrrrrr}
~~~~~&$W^2$ & $Q^2$ & $\Delta\sigma$~~~~~ & $\Delta\sigma_R$~ & $g_1$~~~~~~~~~~~~~~~\\
\hline
$4.5^\circ$&1.31&0.55&  --74$\pm$133&--714&--0.013$\pm$0.023$\pm$0.023\\
&1.69&0.54& --289$\pm$116&--244&--0.070$\pm$0.028$\pm$0.006\\
&2.06&0.53&  627$\pm$106& 282& 0.197$\pm$0.033$\pm$0.039\\
&2.44&0.52&  828$\pm$109& 229& 0.320$\pm$0.042$\pm$0.046\\
&2.81&0.50&  245$\pm$108&--120& 0.113$\pm$0.050$\pm$0.025\\
&3.19&0.49&  120$\pm$~80&--166& 0.064$\pm$0.043$\pm$0.034\\
&3.56&0.48&   75$\pm$~78& --59& 0.046$\pm$0.048$\pm$0.015\\
&3.94&0.47&   --2$\pm$~79& --29&--0.001$\pm$0.055$\pm$0.010\\
&4.31&0.46&  158$\pm$~68&  --7& 0.123$\pm$0.053$\pm$0.015\\
&4.69&0.45&  124$\pm$~66& --10& 0.106$\pm$0.057$\pm$0.010\\
\hline
$7^\circ$&1.19&1.28&   --4$\pm$~5& --19&--0.003$\pm$0.003$\pm$0.003\\
&1.56&1.26&   --8$\pm$16& --49&--0.007$\pm$0.014$\pm$0.005\\
&1.94&1.23&   43$\pm$12&   4& 0.043$\pm$0.012$\pm$0.015\\
&2.31&1.20&  162$\pm$16&  45& 0.193$\pm$0.020$\pm$0.031\\
&2.69&1.18&  121$\pm$16&   3& 0.165$\pm$0.022$\pm$0.016\\
&3.06&1.15&   95$\pm$15& --12& 0.147$\pm$0.024$\pm$0.019\\
&3.44&1.12&   57$\pm$15&   1& 0.098$\pm$0.026$\pm$0.009\\
&3.81&1.09&   70$\pm$15&   2& 0.134$\pm$0.029$\pm$0.012\\
&4.19&1.07&  110$\pm$15&   5& 0.233$\pm$0.031$\pm$0.025\\
&4.56&1.04&   83$\pm$14&   4& 0.193$\pm$0.033$\pm$0.020\\
&4.94&1.01&   63$\pm$14&   3& 0.158$\pm$0.036$\pm$0.022\\
\end{tabular}
\end{table}
\eject

\begin{table}[t]
\caption{Same as Table 2, but for the deuteron.}
\begin{tabular}{rrrrrr}
~~~~~&$W^2$ & $Q^2$ & $\Delta\sigma$~~~~~ & $\Delta\sigma_R$~ & $g_1$~~~~~~~~~~~~~~~\\
\hline
$4.5^\circ$&1.31&0.55& -195$\pm$207&-397&--0.034$\pm$0.036$\pm$0.021\\
&1.69&0.54& --190$\pm$185&--211&--0.046$\pm$0.045$\pm$0.015\\
&2.06&0.53&  401$\pm$163&  80& 0.126$\pm$0.051$\pm$0.016\\
&2.44&0.52&  127$\pm$165&  43& 0.049$\pm$0.064$\pm$0.007\\
&2.81&0.50&  --41$\pm$145& --34&--0.019$\pm$0.067$\pm$0.005\\
&3.19&0.49&  166$\pm$128& --32& 0.089$\pm$0.069$\pm$0.013\\
&3.56&0.48&   36$\pm$120& --17& 0.022$\pm$0.074$\pm$0.004\\
&3.94&0.47&   --8$\pm$112& --14&--0.006$\pm$0.078$\pm$0.002\\
&4.31&0.46&  --27$\pm$101&  --9&--0.021$\pm$0.078$\pm$0.003\\
&4.69&0.45&   67$\pm$97 &  --9& 0.058$\pm$0.084$\pm$0.007\\
\hline
$7^\circ$&1.19&1.28&   79$\pm$24&  18& 0.053$\pm$0.016$\pm$0.010\\
&1.56&1.26&    1$\pm$21& --30& 0.001$\pm$0.018$\pm$0.007\\
&1.94&1.23&   30$\pm$19&   1& 0.030$\pm$0.019$\pm$0.007\\
&2.31&1.20&   26$\pm$22&  11& 0.031$\pm$0.026$\pm$0.005\\
&2.69&1.18&   28$\pm$22&   0& 0.038$\pm$0.030$\pm$0.004\\
&3.06&1.15&   72$\pm$22&   0& 0.112$\pm$0.034$\pm$0.013\\
&3.44&1.12&   19$\pm$22&   3& 0.034$\pm$0.038$\pm$0.004\\
&3.81&1.09&   27$\pm$21&   2& 0.053$\pm$0.041$\pm$0.006\\
&4.19&1.07&   42$\pm$21&   2& 0.089$\pm$0.044$\pm$0.010\\
&4.56&1.04&   34$\pm$21&   1& 0.080$\pm$0.048$\pm$0.008\\
&4.94&1.01&   50$\pm$20&   1& 0.126$\pm$0.051$\pm$0.015\\
\end{tabular}
\end{table}

\eject

\widetext

\onecolumn
\begin{figure}[t]
     \includegraphics{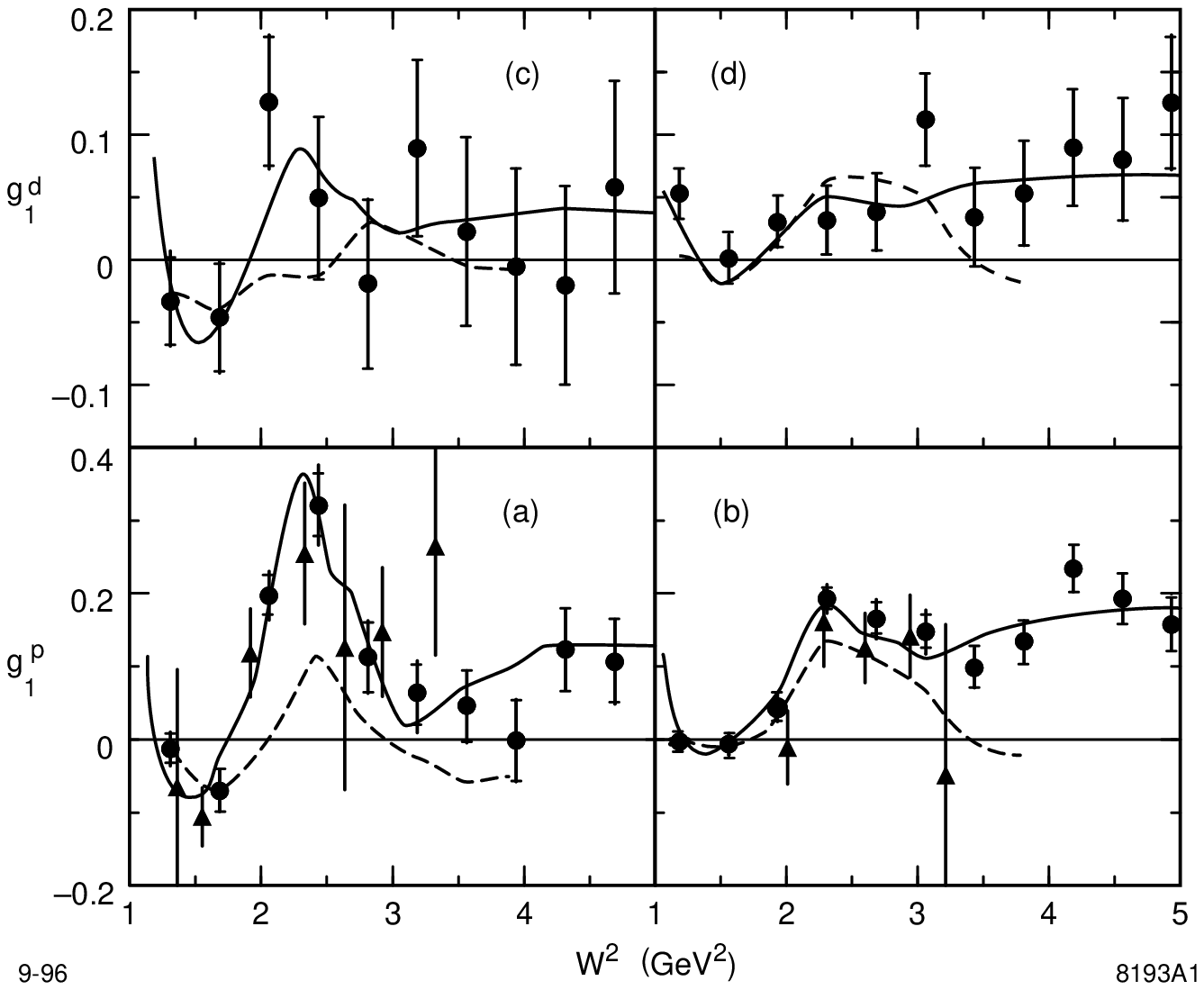}
\vspace*{5.0in}

\end{figure}
\noindent
Fig.~1. Measurements of $g_1$ as a function of $W^2$ for the proton at  (a) $4.5^\circ$
and (b) $7^\circ$; and for the deuteron at (c)  $4.5^\circ$
and (d) $7^\circ$.  The present data (circles) are plotted together with the data of
Baum {\it et al.} (triangles), our Monte Carlo simulation (solid line), and the
model AO of Burkert and Li\protect{\cite{burkert}} (dashed line).
The full error bars correspond to statistical  and
systematic errors added in quadrature, whereas the cross bars indicate
statistical errors only.
\eject

\narrowtext
\begin{figure}[t]
     \includegraphics{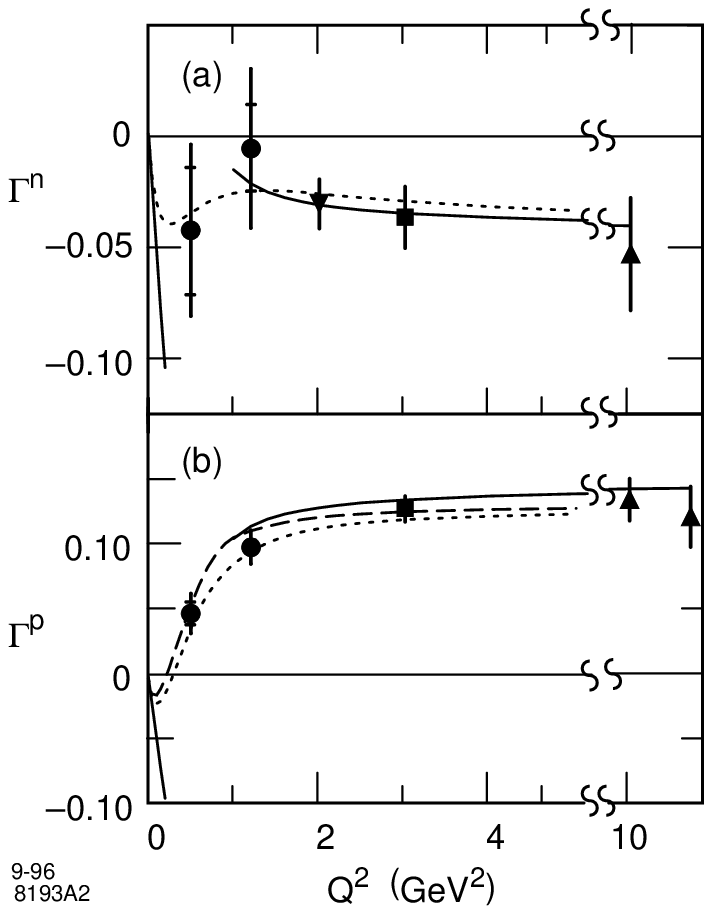}
\vspace*{5.0in}

\end{figure}
\noindent
Fig.~2. Integrals of $g_1$ at several fixed values of $Q^2$ for (a) the neutron
and (b) the proton.  The present data (circles) are plotted together with data from
CERN\protect{\cite{SMC}} (triangles), E143 deep-inelastic\protect{\cite{E143}}
(squares), and E142\protect{\cite{E142}} (inverted triangle).
The curves correspond to the
evolution\protect{\cite{larin}}
of the deep-inelastic results due to changing $\alpha_s$ (solid line),
the predictions of Burkert and Ioffe\protect{\cite{BuIo} (dotted line),
the model of Soffer\protect{\cite{soffer}} (dashed line),
and the GDH approach to $Q^2 = 0$ (solid line).
Errors are indicated as in Fig.~1.
\hskip 1cm


\begin{references}
\bibitem[*]{CRT}
Permanent address: Oliver Lodge Lab, University of Liverpool,
Liverpool, U. K.

\bibitem[**]{BONN}
 Permanent address: University of Bonn, D-53113 Bonn, Germany.

\bibitem[\S]{hoibraten}
Permanent address: FFIYM, P.O. Box 25, N-2007 Kjeller, Norway.

\bibitem[\ddag]{KSU}
Present address: Kent State University, Kent, Ohio 44242.

\bibitem[\dag\dag]{CERN}
Permanent address: CERN, 1211 Geneva 23, Switzerland.

\bibitem[\dag]{raue}
Permanent address: Florida International University, Miami, FL 33199.

\bibitem[\ddag\ddag]{caltech}
Present Address: California Institute of Technology, Pasadena, CA 91125

\bibitem{E80}
E80, M.J.~Alguard, {\it et al.,} Phys. Rev. Lett. {\bf 37}, 1261 (1976);
Phys. Rev. Lett. {\bf 41}, 70 (1978); E130, G. Baum,
{\it et al.,} Phys. Rev. Lett. {\bf 51}, 1135 (1983);

\bibitem{EMC}
EMC, J.~Ashman {\it et al.,} Phys. Lett. {\bf B206}, 364 (1988);
Nucl. Phys. {\bf B328}, 1 (1989).

\bibitem{E142}
E142, P.~L.~Anthony {\it et al.,} Phys. Rev. Lett. {\bf 71}, 959
(1993).

\bibitem{SMC}
SMC, B. Adeva {\it et al.,} Phys. Lett. {\bf B302}, 533 (1993);
D. Adams {\it et al.}, Phys. Lett. {\bf B329}, 399 (1994);
{\bf B336}, 125 (1994); {\bf B357}, 248 (1995).

\bibitem{E143}
E143, K. Abe {\it et al.}, Phys. Rev. Lett. {\bf 74}, 346 (1995);
Phys. Rev. Lett. {\bf 75}, 25 (1995);
Phys. Rev. Lett. {\bf 76}, 587 (1996).

\bibitem{e143q}
E143, K.~Abe {\it et al.,} Phys. Lett. {\bf B364}, 61 (1995).

\bibitem{roberts}
R.~D.~Roberts, The Structure of the Proton, Cambridge Univ.\ Press (1990).

\bibitem{GDH}
S. Gerasimov, Sov. J. Nucl. Phys. {\bf 2}, 430 (1966);
S.D.~Drell and A.C.~Hearn, Phys. Rev. Lett. {\bf 16}, 908 (1966).

\bibitem{keppel}
C. Keppel, PhD Thesis, American University, 1994; unpublished.


\bibitem{wand}
S. Wandzura and F. Wilczek,  Phys. Lett. {\bf  B72}, 195 (1977).

\bibitem{raines}
P. Raines, PhD Thesis, University of Pennsylviania, 1996; unpublished.

\bibitem{stuart}
L.M.~Stuart, {\it et al.}, SLAC-PUB-7391 (1996), HEP-PH 9612416; submitted
to Phys. Rev. D.

\bibitem{radcor} T.~V.~Kukhto and N.~M.~Shumeiko, Nucl.
Phys. {\bf B219}, 412 (1983); I.~V.~Akusevich and N.~M.~Shumeiko,
J. Phys. {\bf G20}, 513 (1994).

\bibitem{baum}
G.~Baum {\it et al.}, Phys. Rev. Lett. {\bf 45}, 2000 (1980).

\bibitem{burkert}
V.~Burkert and Z-J.~Li, Phys. Rev. D {\bf 47}, 46 (1993).

\bibitem{BKM}
V.~Bernard, N.~Kaiser and U.-G.~Meissner, Phys. Rev. D {\bf 48}, 3062 (1993).

\bibitem{Ji}
X.~Ji and P.~Unrau, Phys. Lett. {\bf B333}, 228 (1994).

\bibitem{LiLi}
Z.-P.~Li and Zh.~Li, Phys. Rev. D {\bf 50}, 3119 (1994).

\bibitem{BuIo}
V.D.~Burkert and B.L.~Ioffe, Phys. Lett. {\bf B296} 223, (1992);
CEBAF Preprint PR-93-034.

\bibitem{soffer}
J.~Soffer and O.V.~Teryaev, Phys. Rev. D {\bf 51}, 25 (1995).

\bibitem{larin}
S.A.~Larin, Phys. Lett. {\bf B334}, 192 (1994).

\bibitem{buco}
H. Burkhardt and W. N. Cottingham,  Ann. Phys. {\bf  56}, 453  (1970).

%
%
%
%
%
%
%
%
%
%
%
%
%
%
%
%
%
%
%
%
%

\end{references}
\end{document}